\documentclass[runningheads]{llncs}
\usepackage{graphicx}
\usepackage{xcolor}
\usepackage{float}
\usepackage{amsmath}
\usepackage{soul}
\usepackage[symbol]{footmisc}

\begin{document}
\title{A for-loop is all you need. For solving the inverse problem in the case of personalized tumor growth modeling}
\author{Ivan Ezhov\inst{1,2} \and Marcel Rosier\inst{1} \and Lucas Zimmer\inst{2,3} \and Florian Kofler\inst{1,2,4} \and Suprosanna Shit\inst{1,2} \and Johannes Paetzold\inst{1,2} \and Kevin Scibilia\inst{1} \and Leon Maechler\inst{5} \and Katharina Franitza\inst{1} \and Tamaz Amiranashvili\inst{1,3,6} \and Martin J. Menten\inst{1,2} \and Marie Metz\inst{2,4} \and Sailesh Conjeti\inst{7} \and Benedikt Wiestler\inst{2,4,*} \and Bjoern Menze\inst{3,}\thanks{Contributed equally as senior authors}
}
\institute{Department of Informatics, TUM, Munich \and TranslaTUM - Central Institute for Translational Cancer Research, TUM, Munich \and Department of Quantitative Biomedicine, UZH, Zurich \and Neuroradiology Department of Klinikum Rechts der Isar, TUM, Munich \and École Normale Supérieure, Paris \and Visual and Data-Centric Computing, Zuse Institute Berlin, Berlin \and Siemens Healthineers, Germany\\ \email{ivan.ezhov@tum.de}}

\maketitle              
\begin{abstract}
Solving the inverse problem is the key step in evaluating the capacity of a physical model to describe real phenomena. In medical image computing, it aligns with the classical theme of image-based model personalization. Traditionally, a solution to the problem is obtained by performing either sampling or variational inference based methods. Both approaches aim to identify a set of free physical model parameters that results in a simulation best matching an empirical observation. When applied to brain tumor modeling, one of the instances of image-based model personalization in medical image computing, the overarching drawback of the methods is the time complexity for finding such a set. In a clinical setting with limited time between imaging and diagnosis or even intervention, this time complexity may prove critical. 
As the history of quantitative science is the history of compression \cite{juergen}, we align in this paper with the historical tendency and propose a method compressing complex traditional strategies for solving an inverse problem into a simple database query task. We evaluated different ways of performing the database query task assessing the trade-off between accuracy and execution time. On the exemplary task of brain tumor growth modeling, we prove that the proposed method achieves one order speed-up compared to existing approaches for solving the inverse problem. The resulting compute time offers critical means for relying on more complex and, hence, realistic models, for integrating image preprocessing and inverse modeling even deeper, or for implementing the current model into a clinical workflow.

\end{abstract}

\section{Introduction}

Magnetic resonance imaging (MRI) is the gold standard technique to diagnose brain tumors, such as glioblastoma (GBM). While being able to reliably highlight areas of sufficiently high tumor cell concentration in GBMs, it can lack the capacity to visualize areas of lower tumor cell density at the tumor border and most importantly areas of diffuse tumor infiltration, a key biological property of GBM. Current radiotherapy plans try to account for the unknown infiltration by targeting a uniformly extended volume around the tumor outlines visible in MRI. While decreasing the probability of tumor recurrence, such a treatment planning has an obvious drawback of unnecessarily damaging healthy tissue, which in turn has a negative impact on the patient’s life quality. Personalizing the target of radiotherapy by complementing the MRI scans with individual tumor simulation, that models a complete spatial distribution of tumor cell concentration, could preserve healthy tissue and reduce the risk for secondary malignancies and side effects \cite{Le_tmi,le_miccai,jana_tmi,hormuth2021image}.

Conventional approaches for simulation-based personalization attempt to model the tumor growth for each individual patient using differential equation solvers. The personlaization is achieved by solving the inverse problem - identifying tumor model's parameters best matching tumor signal from MRI. However, utilizing the numerical solvers for solving an inverse problem still results in extreme runtimes which obstruct transfer into clinical practice. To address this issue, highly efficient model- and data-driven approaches were developed over the recent years \cite{subramanian2020multiatlas,scheufele2019coupling,hormuth2018mechanically,hormuth2021image,jana_tmi,petersen2019deep,petersen2021continuous}. The time for solving an inverse problem using numerical model-based solvers hit one hour of compute \cite{subramanian2020multiatlas,scheufele2019coupling,hormuth2021image}. For data-driven approaches the computing time can be reduced to minutes \cite{petersen2019deep,petersen2021continuous,Ezhov_2019,Ezhov2020GeometryawareNS}, but they all rely on some type of a neural network to predict tumor progression. Neural networks are known to be non-robust when extrapolated to out-of-training distributions - the key obstacle in integrating such algorithms into safety-critical applications. 

In this paper, we propose an image retrieval based approach that performs a query of
patient specific scans to a database of synthetic tumors, returning the closest resemblance of the patient’s tumor. As a baseline, this image retrieval process is implemented via a primitive iterative pair-wise comparison. Further, we investigate how a retrieval in low dimensional embeddings of the simulations -- that we obtain by using downsampling as well as autoencoders, variational autoencoders, unsupervised hashing, and radiomics feature representation -- can improve runtimes. As a result, our work shows that our query approach can yield accurate and, depending on the chosen optimization, also deterministic results in the order of seconds (correspondingly, minutes for the inverse problem). 

\begin{figure}[h]
\centering
\includegraphics[width=1.0\textwidth]{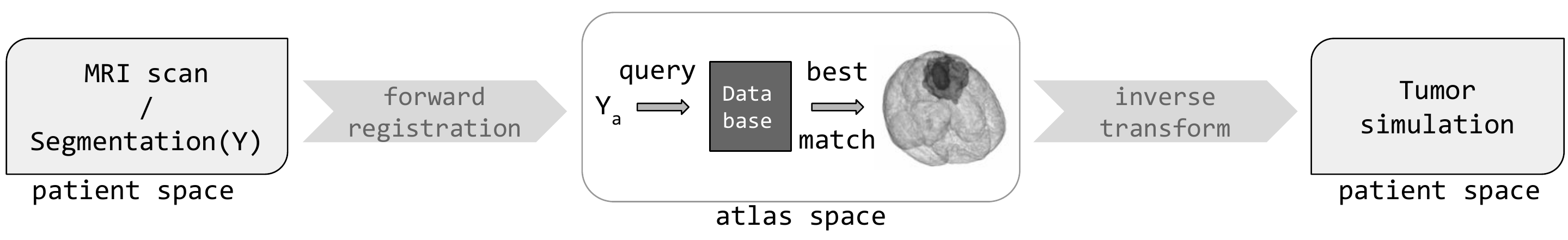}
\caption{\small{Schematic illustration of the proposed pipeline. First, we register a patient MRI scan to the brain atlas. Next, we use the obtained forward transformation matrix to morph the patient segmentation to the atlas space. Then, we perform a query to a database of 50k pre-simulated tumors in the atlas space. Essentially, in this step we find the closest to the query sample binary volume in the database. Finally, we transform a tumor simulation, corresponding to the best matching binary volume, back to the patient space.}}
\label{qualcomp}
\end{figure}
\section{Method}
\subsection{Tumor growth model.}

Our forward model is based on the most used within the medical imaging community mathematical formalism - the reaction-diffusion partial differential equation. The equation describes the evolution of tumor cell density through proliferation and diffusion of the cells:   

\begin{equation}
\frac{\partial u}{\partial t}= \nabla (\textbf{D}\nabla u)+\rho u(1-u), \quad in \quad\Omega \\
\end{equation}
\begin{equation}
\textbf{n} \cdot \nabla u = 0, \quad in \quad\Gamma_{\Omega}.
\end{equation}

Here, the first term on the right-hand site of the equation denotes the diffusion process, where \textbf{D} is the diffusion tensor. The second term models the logistic proliferation of cells, and $\rho$ denotes the rate of cell proliferation. We impose the Neumann boundary condition on the boundary of the simulation domain $\Gamma_{\Omega}$, while $\Omega$ denotes the simulation domain itself, i.e. volume of the brain, and \textbf{n} is the unit vector orthogonal to $\Gamma_{\Omega}$.
As an initial condition for the cell density $u$ we use a seed point at a spatial location $\pmb{r^*}$: $u(\pmb{r},0)=u_0$ if $\pmb{r}=\pmb{r^*}$, $u(\pmb{r},0)=0$ if $\pmb{r}\neq\pmb{r^*}$. The solver outputs a simulated tumor density $u$ of size $128^3$.

We assume that visible outlines in an MRI image correspond to isolines in $u$, and to link the tumor simulations with an imaging signal, we threshold the simulations at $u_{T1Gd}, u_{FLAIR}$. This allows us to reproduce the binary segmentation obtained from an MRI scan (T1Gd, FLAIR) delineation \cite{le_miccai,Ezhov_2019}.

\subsection{Inverse model.}

The proposed inverse model pipeline consists of the following steps, Fig. 1: 

1. A patient MRI scan is registered to the
atlas brain anatomy \cite{rohlfing2010sri24}, yielding a transformation matrix $M$. 

2. $M$ is used to morph the corresponding segmentation $Y$ into the atlas space resulting in $Y_a$. 

3. A query is made to a database of synthetic tumors which were simulated beforehand in the brain atlas. This query finds the closest match\footnote[1]{The closest match can be determined in different ways which we are comparing in this study} to $Y_a$, and returns the corresponding tumor simulation in the atlas space.

4. As a last step, utilizing the inverse of $M$, the atlas simulation is morphed back to the patient's brain anatomy to obtain the patient-specific simulation.

\subsection{Query to a database of synthetic tumors. }
The input MRI scans consist of two segmentations while the synthetic tumors are a spatial distribution containing continuous values in the range from 0 to 1. Therefore, we have to extract the desired segmentations from the raw synthetic tumor data. For this purpose, we threshold the
synthetic tumors at $u_{FLAIR}=0.2$ to receive the FLAIR and at $u_{T1Gd}=0.6$ for the T1Gd segmentations \cite{Le_tmi,martens2021deep}. 

After preprocessing the dataset to make it comparable, we need to select a fitting metric. The Dice coefficient is a common measure for medical imaging segmentations, we therefore utilized it as the main metric. 

Even though we believe that a for-loop is all you need, in order to please reviewers spoiled with deep learning sugar, we evaluated different learnable strategies to perform a query to a database of simulated tumors. For some of these query strategies, in order to evaluate similarities between the compressed vectors, we use the L2 = $\left\| \cdot \right\|_2 $ metric, as Dice is barely meaningful e.g. for comparing latent representations. Furthermore, since we have to compare two segmentations for each tumor, it is possible that the separate queries result in different best matches. Hence, the two query results have to be combined to determine a joint best match. For simplicity, we decided to simply add the two individual similarity values and choose the highest combined value for Dice ($Dice_{T1Gd}+Dice_{FLAIR}$) or the lowest for L2 ($L2_{T1Gd}+L2_{FLAIR}$) as the best match. 

\subsubsection{Baseline for-loop search.} The most basic way to implement the query is by naively looping over the dataset and performing a pair-wise comparison between input and database tumors.
We implemented this basic brute-force loop to create a ground truth that serves
as a reference for later improvements and experiments. Since the individual comparisons
are not dependent on each other, we implemented the query by parallelizing the loop to
shorten its runtime.

Although parallelization can help to reduce the time complexity of the database query, the
speed of this operation is still limited by the significant memory usage. This restriction also
prohibits the use of efficient query frameworks. For instance, the FAISS library is designed
for vector sizes in the range of roughly 20 - 2000 \cite{johnson2019billion}. However, a plain flattening of our 3D volume of $128^3$ voxels exceeds this suggested limit by a factor of more than 1000. Therefore, we decided to experiment with a compression of the data.

As a first step, we try a naive downsampling of the volume to lower resolutions. In our case, we tried a down sampling to $64^3$ and $32^3$ voxels. We downsample the original volume by applying a spline interpolation of order 0 to the original volume \cite{virtanen2020scipy}. We then compare downsampled database with a downsampled query volume, and choose the one that has the highest Dice in the downsampled volume.

\subsubsection{Autoencoders.} To obtain even more compact data representation, we experimented with convolutional autoencoders (AE) and variational autoencoders (VAE). We examined if these techniques are capable of creating a problem-specific encoding that might be able to sufficiently preserve similarity relationships between tumors, while drastically reducing the runtime. The core idea for our use-case is to train the AE to extract meaningful properties of the tumors and then apply the encoder function to the dataset $\{Y\}$, yielding an encoded dataset $\{Y_{enc}\}$. For each query, we utilize $\{Y_{enc}\}$ and encode the input tumor using the same encoder, resulting in a data size drop from $128^3$ to the dimension of the latent space (1024). Since the segmentations of T1Gd and FLAIR scans differ in their properties, two separate AEs are needed, leading to two encoded datasets $\{Y_{enc}^{T1Gd}\}$ and $\{Y_{enc}^{FLAIR}\}$.

To further improve the performance of our AEs, we experimented with VAEs. Basic AEs have the potential disadvantage of limited control over the latent space. VAEs enforce a distribution over the latent space by penalizing deviations from a previously selected distribution during training, which can produce more continuous and organized latent spaces while still providing adequate reconstructions \cite{kingma2013auto}. Usually, this property is used to generate previously unseen output by randomly sampling from the latent space. However, we investigated if this characteristic is also beneficial for similarity preservation. 

In the end, we compare latent representation of the database with a latent representation of the query volume, and choose the one that has the lowest L2.

\subsubsection{Unsupervised hashing.} An alternative compression strategy is to use a hashing - compression to a bit string representation \cite{hashing1,hashing2,hashing3,hashing4}. We resorted to the most recent deep unsupervised hashing technique \cite{liAAAI2021} for our experiments. The methodological idea is a replica of the AE with a difference in how the the latent space is formed. In \cite{liAAAI2021}, the continuous output of the AE's encoder is quantized via a so-called "Bi-half" hash layer into a binary representation. This representation serves then as an input to the AE's decoder. All other network elements are identical to the conventional AE setup. Similarly to the AE and VAE, we choose the best match that has the lowest L2 when compared to the binary representation of the query sample.

\subsubsection{Radiomics shape feature representation.} The ultimate compression strategy is to use single value characteristics of a 3D shape and morphology. We experimented with shape features from the pyradiomics library \cite{van2017computational}, namely voxel volume, elongation, major axis length, etc. Additionally, we computed the tumor's center of mass to account for its location. For the query, we used the following strategy. We selected 1000 samples from our database which have the closest center of mass to our query sample in the atlas reference space. Out of these 1000 samples, we selected the one that is closest to the query sample under L2 measure evaluated for the shape features: $\sum_i \left\| f_q^i, f_d^i \right\|_2$. Here, the summation is over the radiomics features, $f_q^i$ is the feature representation computed for the query sample, and $f_d^i$ is the feature computed for the database sample.

\subsection{Implementation.}


The base architecture for our AE is a convolutional neural network. In the encoder, we scale down
the volume layer-by-layer using strided 3D convolutions (stride=2, kernel size=3) followed by
a convolution without stride, while increasing the number of channels. After downscaling
the volume, we flatten the features and apply a linear layer to obtain a latent representation of size 1024. Afterwards, the decoder mirrors this process with strided 3D transposed convolutions. We executed multiple test runs searching for the best architecture that suggested that a downscaling to a resolution of $16^3$ before flattening the volume and 1024 as the size of the latent representation yield the most reasonable results. We use this resolution throughout all consecutive experiments. For comparability, our VAE network replicates the AE network (except the linear layer before the latent space that is replaced by 2 separate linear layers which output the mean and the logarithm of the variance). AE, VAE, and the hashing method were trained on a dataset of 1500 tumors (larger training set sizes showed no significant improvements) along with a validation set of size 150. 

We used the Advanced Normalization Tools (ANTs) \cite{avants2009advanced} for the registration. with basic settings. For the looping-based queries, we performed the compute on AMD EPYC 7452 CPU and parallelized the search over 32 processes. For AE, VAE, and BF-HASH, the training and evaluation were performed on an   NVIDIA QUADRO P8000 GPU. 

\section{Experiments}
\subsubsection{Data.}
The basis for our work is a dataset of 50,000 synthetic tumor volumes of $128^3$ resolution. In order to generate synthetic tumors, a common brain anatomy is required that is able to represent a broad range of realistic human brains. For this purpose, the atlas brain anatomy introduced in \cite{rohlfing2010sri24} was utilized. This represents a statistical average for the relevant distributions of cerebrospinal fluid (CSF), white matter (WM) and gray matter (GM), combined into a single atlas space. In this atlas space, a tumor can then be simulated using the reaction diffusion model by setting values for the initial location $ x, y, z $ , the proliferation rate $\rho$, the diffusion rate in WM $D_w$ as well as the simulation end time $T_{end}$, which corresponds to the tumor’s age. The synthetic dataset was generated by randomly sampling patient-specific parameters from the subsequent ranges and feeding the resulting parameter sets
$\theta = \{ x, y, z, D_w , \rho, T_{end} \}$ to the reaction diffusion model. Analogous to \cite{ezhov2021learn}, we discarded unrealistic in size tumors based on the range of real tumor sizes (BraTS dataset \cite{menzebrats}).

In all experiments, we evaluate the inverse model pipeline on 62 real MRI scans of patients diagnosed with gliomas for which T1Gd and FLAIR MRI scanning were performed. For these patients, the corresponding tumor annotations were obtained by expert radiologists' labeling. In addition, for evaluation on synthetic data, we used 1k tumors simulated in the atlas space but not present in the query database. 

\begin{table}
  \centering
  \begin{tabular}{ |c||c|c|c|c|c|c|c| }
         \hline
          & direct query & DS 64 & DS 32 & VAE & AE & BF-HASH &  RF\\
         \hline
         Top-1              & 100\% & 88.7\% & 66.1\% & 41.9\% & 33.8\%  & 32.2\% & 4.92\% \\
         Top-5              & 100\% & 100\% & 96.7\% & 61.2\% & 64.5\%  & 67.7\% & 4.92\%\\
         Top-15             & 100\% & 100\% & 100\% & 80.6\% & 77.4\%  & 82.2\% & 9.83\% \\
         \hline
         Runtime     & 340s & 130s & 116s & 34s & 22s & 22s & 2.8s\\
         \hline
  \end{tabular}
  \centering
  \caption{Comparison between different best match query strategies: looping over the dataset with resolution $128^3$ (direct query), looping over downsampled to $64^3$ dataset (DS 64), looping over downsampled to $32^3$ dataset (DS 32), using embeddings obtained by the VAE and AE, using the radiomics features (RF), and using hashing (BF-HASH). Top N accuracy represents a measure whether the predicted best match falls in the top N closest predictions. For the query using direct, DS64, and DS32 methods the Dice score is used, while for the remaining methods we use the $L2$ measure. We expectedly observe that an improvement in runtime comes at a cost of query accuracy.}
\end{table}

\begin{table}
 \centering
 \begin{tabular}{c|c|c|c||c|c|c}

        \hline
        & \multicolumn{3}{c}{Real} &  \multicolumn{3}{c}{Synthetic} \\
        \cline{2-7}
         & Top-1 & Top-5 & Top-15  & Top-1 & Top-5 & Top-15 \\
        \hline
        direct query& 100\% & 100\% & 100\%  & 100\% & 100\% & 100\% \\
        \hline
        DS64 & 88.7\% & 100\% & 100\%   & 93.5\% & 100\% & 100\% \\
        \hline
        DS32 & 66.1\% & 96.7\% & 100\%  & 78.9\% & 99.6\% & 100\% \\

        \hline
 \end{tabular}
 \centering
 \caption{Comparison between different loop-based query strategies on real (62 cases) and synthetic (1000 cases): looping over the dataset with resolution $128^3$ (direct query), looping over downsampled to $64^3$ dataset (DS 64), looping over downsampled to $32^3$ dataset (DS 32). Top-N accuracy represents a measure whether the predicted best match falls in the top N closest predictions. The average DICE (sum of DICEs for T1Gd and FLAIR) for best matching samples is 1.07 for real, and 1.69 for synthetic samples.}
\end{table}

\subsubsection{Quantitative comparison between query methods.} Table 1 compares the presented database query approaches on the real dataset: 1) simple looping over the dataset with original resolution $128^3$ (direct query), 2) looping over downsampled to $64^3$ dataset (DS 64), 3) looping over downsampled to $32^3$ dataset (DS 32), 4) using embeddings obtained by the VAE and AE, 5) using the radiomics features (RF), and 6) and unsupervised hashing (BF-HASH). The top-N accuracy is computed with respect to the best match samples found by the baseline. As expected, we observe that an improvement in runtime comes at a cost of losing accuracy of similarity preservation.

\begin{figure}[ht]
\centering
\includegraphics[width=1.0\textwidth]{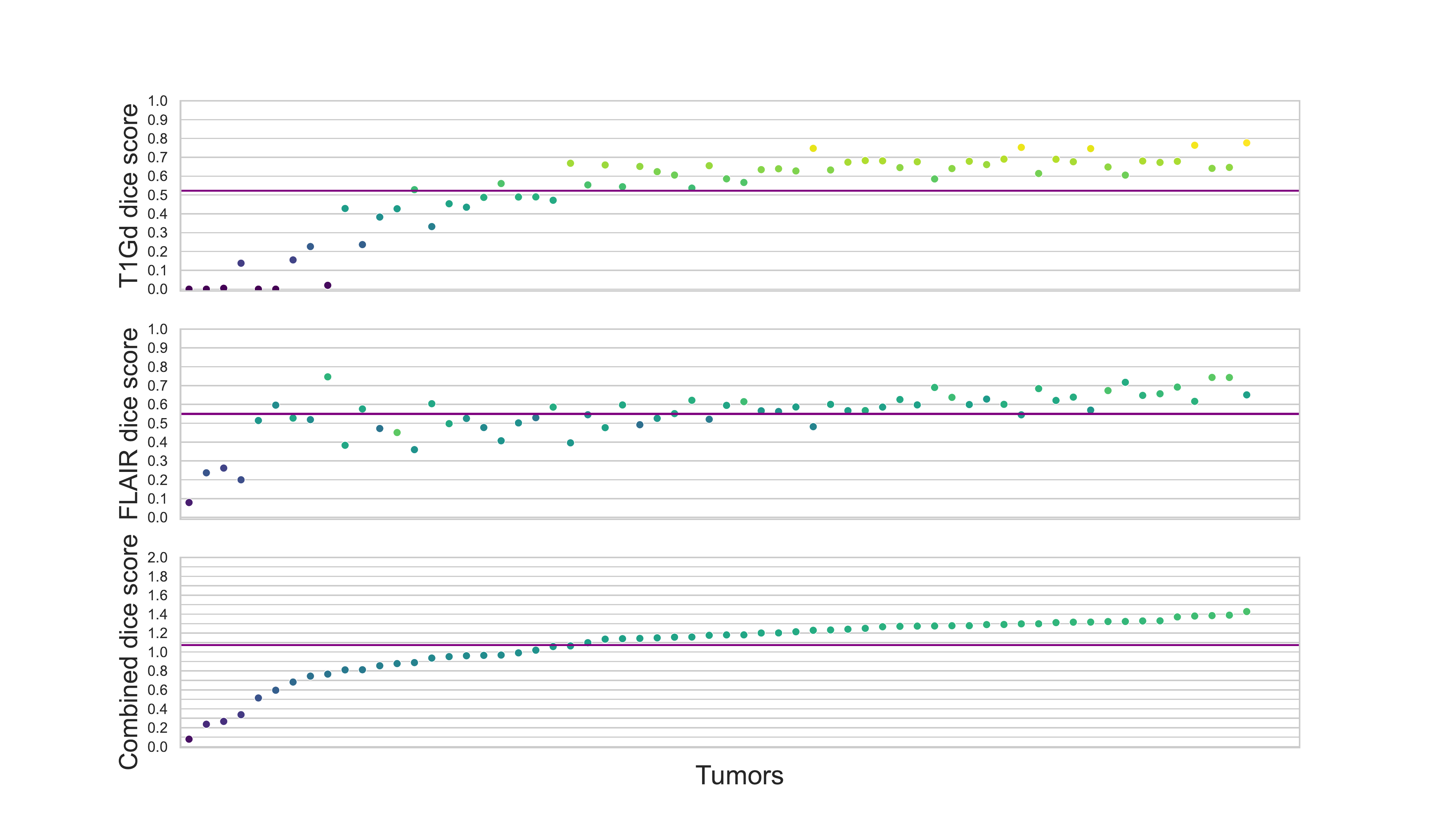}
\caption{\small{DICE between real tumor segmentation (62 patients) and best matching synthetic tumor in the atlas space. DICE is computed for T1Gd (upper row), FLAIR (middle row), and combined T1Gd+FLAIR (down row).}}
\label{distr_real}
\end{figure}

\begin{figure}[h]
\centering
\includegraphics[width=1.0\textwidth]{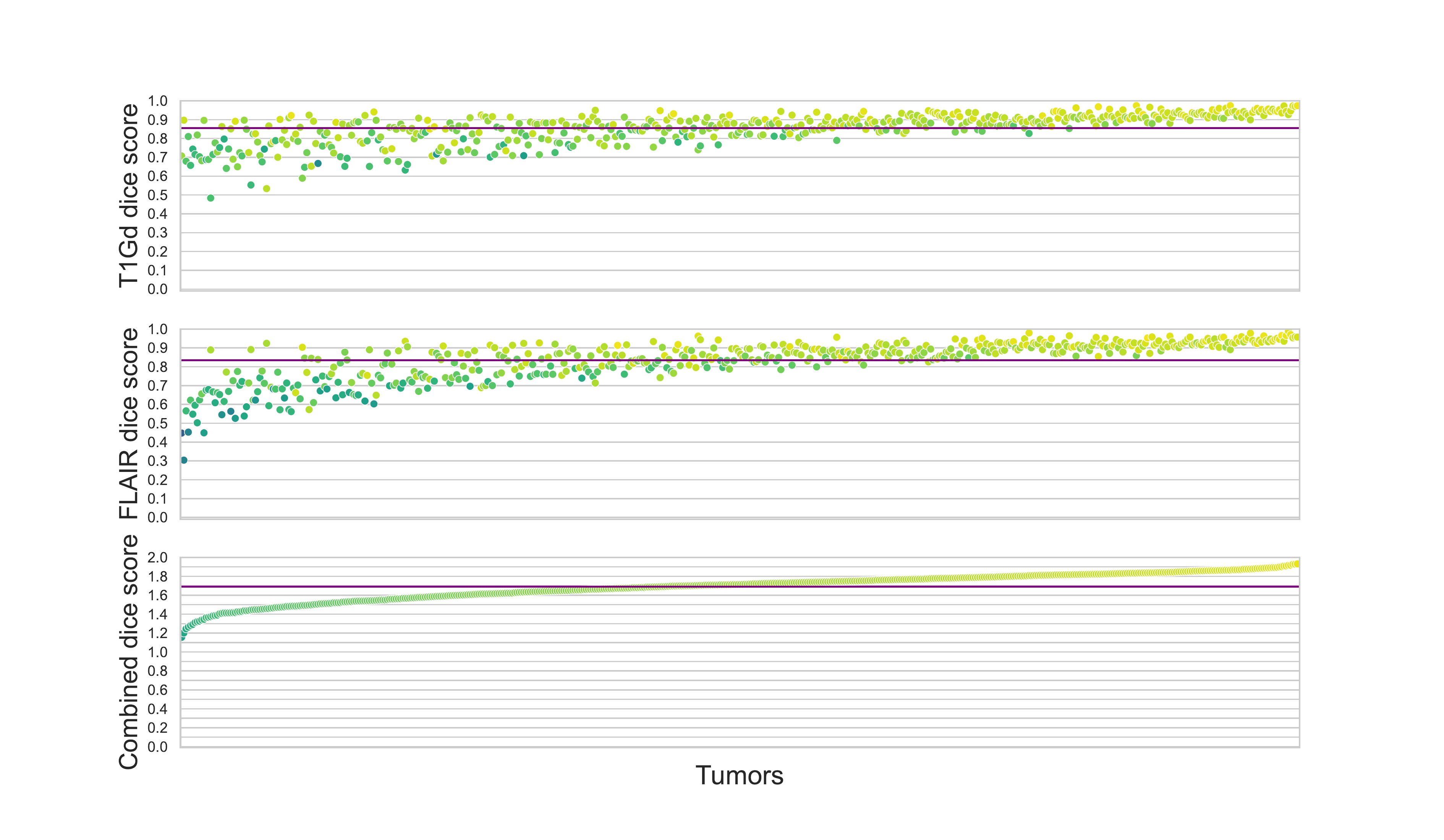}
\caption{\small{DICE between synthetic tumor segmentation (1000 examples) and best matching synthetic tumor in the atlas space. DICE is computed for T1Gd (upper row), FLAIR (middle row), and combined T1Gd+FLAIR (down row).}}
\label{distr_syn}
\end{figure}

Tab.2 showcases how the difference between real and synthetic data distributions affects the query accuracy. Given there is an evident gap between real tumor progression and the tumor growth trajectory modeled by the reaction-diffusion model, we expectedly observe ca. 10 percent drop of the accuracy. Fig. \ref{distr_real} and \ref{distr_syn} demonstrate the same difference in real and synthetic data distributions for every example used in the study.
In a way, in future works probing more complicated tumor descriptions, such analysis can serve as a measure of plausibility of a tumor growth model.


\begin{figure}[hb]
\centering
\includegraphics[width=1.0\textwidth]{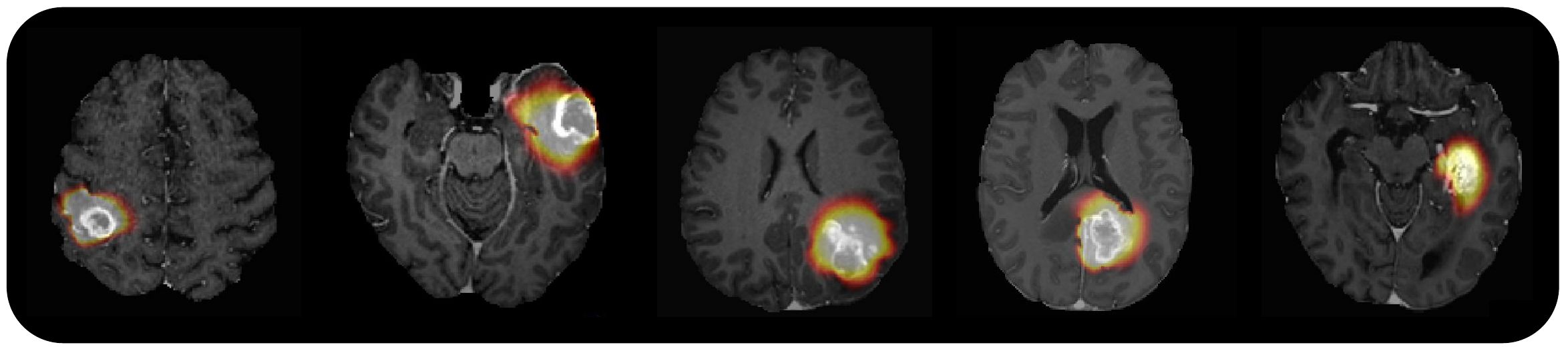}
\caption{\small{Qualitative examples of the tumor simulations obtained from our proposed pipeline. The brain images correspond to five different patients.}}
\label{qualcomp}
\end{figure}

\section{Discussion}

Here we want to discuss the pros and cons of the proposed method. First, the method is deterministic - it relies on direct computation of the best match among tens of thousands of precomputed simulations under the measure of interest. Even though we provided a comparison with different query strategies analysing the trade-off between accuracy and time, by using the direct query strategy one obtains "global optimum" for the query at an acceptable runtime, so we suggest to use it in practice. Also despite using the DICE score as such a measure (since it is the first metric of choice in our community), the method is evidently generalizable to any arbitrary measure. The overall runtimes for the solving the inverse problem, including registration, query, and back transform vary from two (for the inaccurate RF query) to eight minutes (for the optimal direct query). This is an order of magnitude faster than any existing brain tumor inverse problem solving methods (starting from ca. 1 hour for the same resolution by \cite{subramanian2020multiatlas,scheufele2019coupling}). One may question whether such a speed-up comes at the cost of a significant error, which in turn comes from the registration to and finding the best matching in the atlas space. While we would agree with such a statement, it is worth to mention that dominant majority of existing literature resorts to such atlas registration in order to approximate the patient specific anatomy.

\section{Conclusion}

We present a method for inferring a tumor simulation from information available on medical scans relying on a simple database query strategy. A reader might say that the proposed method is a joke due to its offensively simple nature. While we would again agree with such a statement, we do believe that the method provides a generalizable, fast, and robust solution to the inverse problem. The runtimes for solving the inverse problem are in the order of a few minutes which is faster than any existing inverse modeling method. Even though the used database is composed of simulations from the reaction-diffusion model that might not be sufficient to provide a close match to real tumor growth in general, it is evident that the proposed method is generalizable to more sophisticated models (one just needs to resimulate tumor database in the atlas space with more complicated tumor model). As opposed to data-driven approaches, the method does not fail at extrapolation as it is based on a plain compute of the overlap measure between two binary volumes.\\


\bibliographystyle{splncs}
\bibliography{mybib}

\end{document}